\documentclass[%
reprint,
superscriptaddress,
 amsmath,amssymb,
aps, 
prb,
]{revtex4-2}

\usepackage{float}
\usepackage[dvipsnames]{xcolor}

\usepackage{braket}
\usepackage{graphicx}
\usepackage{dcolumn}
\usepackage{bm}
\usepackage[colorlinks=true, allcolors=Blue]{hyperref}

\newcommand{\bk}{\mathbf{k}}
\newcommand{\bG}{\mathbf{G}}
\newcommand{\bQ}{\mathbf{Q}}
\newcommand{\bq}{\mathbf{q}}

\newcommand{\bkbar}{\bar{\bk}}

\begin{document}

\preprint{APS/123-QED}

\title{Various electron crystal phases in rhombohedral graphene multilayers 
}%

\author{Wangqian Miao}
\email{Contact author: wqmiao@psu.edu}
\affiliation{Department of Physics, The Pennsylvania State University, University Park, Pennsylvania 16802, USA}
\affiliation{Center for Theory of Emergent Quantum Matter,
The Pennsylvania State University, University Park, Pennsylvania 16802, USA}
\affiliation{Department of Physics, The Hong Kong University of Science and Technology, Clear Water Bay, Hong Kong, China}

\author{Chu Li}
\affiliation{Department of Physics, The Hong Kong University of Science and Technology, Clear Water Bay, Hong Kong, China}

\date{\today}

\begin{abstract}
We systematically investigate the emergence of electron crystal phases in rhombohedral multilayer graphene using comprehensive self-consistent Hartree Fock calculations combined with \textit{ab initio} tight binding model. As the carrier density increases, we uncover an isospin cascade sequence of phase transitions that gives rise to a rich variety of ordered states, including electron crystal phases with non-zero Chern numbers. We further show the nearly degeneracy of these topological electron crystals hosting extended quantum anomalous Hall effect (EQAH) in the mean field regime and characterize pressure driven phase transitions. Finally, we discuss the thermodynamic signatures, particularly the behavior of the inverse compressibility, in light of recent experimental observations.
\end{abstract}

\maketitle


\section{Introduction}

In the dilute limit of a two dimensional electron gas, kinetic energy is quenched by Coulomb interactions and the electrons spontaneously crystallize into a Wigner crystal. This phenomenon becomes even more intriguing when the underlying electronic bands carry non trivial topology, which may carry integer Chern number \cite{tesanovic_hall_1989_prb, soejima_ensuremathlambda-jellium_2025, desrochers_electronic_2025, Kim_magnetic_2025, joy_chiral_2025, reddy_quantum_2025}.

Experiments on rhombohedral multilayer graphene \cite{lu_fractional_2024,lu_extended_2025, xie_tunable_2025} (RMG) have revealed a quantized Hall effect and even fractional Hall signals upon slight electron doping. The most striking observation is an extended quantum anomalous Hall (QAH) phase that persists in the regime where a strong displacement field pushes the electrons away from the aligned hBN substrate. This suggests that moir\'e effects are relatively weak in this setup, especially when compared with systems such as twisted MoTe$_2$. Recent nano-ARPES studies \cite{zhang_correlated_2024, zhang_moire_2025} indicate that hBN alignment enhances band flatness, in qualitative agreement with quantum capacitance data showing that stronger incompressibility emerges mainly at integer filling factors at moir\`e proximal region \cite{aronson_displacement_2025}. These findings naturally raise the question of the microscopic mechanism behind the insulating phase: is it predominantly driven by electron--electron interactions \cite{dong_anomalous_2024,dong_theory_2024, dong_stability_2024, zhou_fractional_2024_prl, zhou_new_2025, tan_parent_2024, tan_variational_2025, guo_correlation_2025, soejima_anomalous_2024, bernevig_berry_2025, tan_ideal_2025, desrochers_elastic_2025, wei_edge-driven_2025, huang_fractional_2025, aguilar-mendez_full_2025, guo_fractional_2024,kwan_moire_2025}, or does it still originate from residual hBN alignment \cite{uzan_hbn_2025}? From a theoretical perspective, the single-particle Hamiltonian of RMG is relatively simple to investigate: its low-energy physics can be captured by a \(n\)-th order Dirac Hamiltonian \cite{zhang_spontaneous_2011_prl} acting in the layer degree of freedom, effectively describing the first conduction and first valence bands. Remarkably, these bands resemble drumhead surface states \cite{chen_surface_2019_prl} in the bulk limit, which helps to explain both their pronounced flatness in the conduction band bottom and their close analogy to the zeroth Landau level \cite{lau_Designing_2021_prx, tan_ideal_2025}.

\begin{figure}
\includegraphics[width=0.48\textwidth]{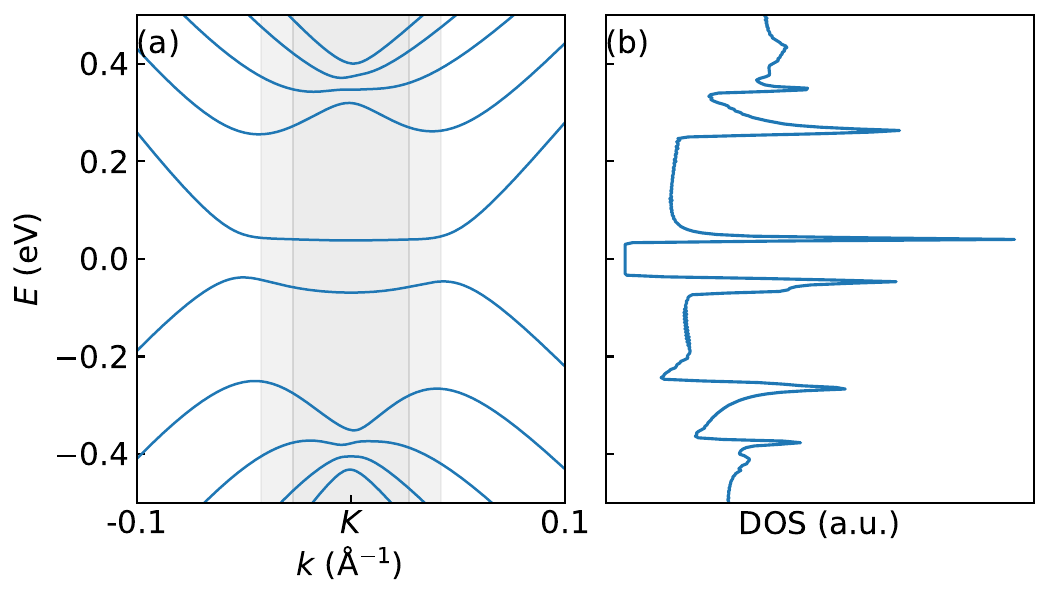}
\caption{\label{fig1}  
Single particle band structure (a) of rhombohedral penta layer graphene (R5G) near the $K$ point for a displacement field of $U = 0.035~\text{eV}$, and the corresponding density of states (b). The van Hove singularity in the first conduction band originates from the flat band bottom. The shaded area lables the region when the electron filling is $0.5\times10^{12}$ cm$^{-2}$ and $1.2\times10^{12}$ cm$^{-2}$, respectively.
}
\end{figure}

In this work, we employ an \emph{ab initio} tight binding model \cite{herzog-arbeitman_moire_2024} for rhombohedral multilayer graphene and systematically map out the Hartree Fock phase diagram as a function of carrier density $n$ and displacement field. Within our mean field framework, we allow for spontaneous breaking of translational symmetry, as well as spin $\mathrm{SU}(2)$ and valley $\mathrm{U}(1)$ symmetries, to capture the onset of electron crystallization. A central outcome is a cascade of isospin phase transitions driven by a Stoner instability \cite{dong_isospin-_2023}, closely analogous to that inferred from inverse-compressibility measurements in rhombohedral trilayer graphene \cite{zhou_half-_2021}. In the same parameter regime, we identify a sequence of electron crystal phases, both topologically trivial and with nonzero Chern number, that energetically outperform the corresponding Fermi-liquid (FL) states at the mean-field level and emerge as stripe-like regions in the phase diagram. Notably, anomalous Hall crystal phases can exist even at relatively high doping, around $n \approx 3 \times 10^{12}\,\mathrm{cm}^{-2}$, at appropriate displacement fields. Motivated by experiments, we then focus on the density range near $n \approx 1 \times 10^{12}\,\mathrm{cm}^{-2}$. In this regime, our numerics reveal strong competition among multiple electronic phases with nearly degenerate energies, suggesting that phase coexistence or mesoscale inhomogeneity may be relevant. This sensitivity also underscores the key role of hBN alignment in stabilizing the QAH state in experiment. Finally, we discuss how moderate hydrostatic pressure can readily tune the WC--AHC transition while largely preserving the favorable quantum geometry of the bands. We conclude by discussing the thermodynamic signatures of these phase transitions, with particular emphasis on the behavior of the inverse compressibility.

\section{Microscopic Hamiltonian}
\subsection{Slater--Koster tight binding model for Rhombohedral multilayer Graphene \label{sec:tb}}

To describe the low-energy electronic structure of rhombohedral multilayer graphene (RMG), we employ a Slater--Koster (SK) parameterization of the $p_z$-orbital tight binding Hamiltonian recently developed in Ref.~\cite{herzog-arbeitman_moire_2024}, benchmarked against density functional theory (DFT). The hopping amplitude between two $p_z$ orbitals separated by vector $\mathbf{r}$ with vertical projection $z = \hat{z}\cdot\mathbf{r}$ is written as
\begin{equation}
    \begin{aligned}
       t_{\text{SK}}(r) = &V_{pp\pi}\!\left(1 - \frac{z^2}{r^2}\right) e^{q_\pi (1 - r/a_\pi)} f_c(r) \\
&+ V_{pp\sigma}\frac{z^2}{r^2} e^{q_\sigma (1 - r/a_\sigma)} f_c(r), 
    \end{aligned}
\end{equation}
where $f_c(r) = \big(1 + e^{(r-r_c)/l_c}\big)^{-1}$ is a smooth cutoff function in the real space. 
The optimized parameters fitted from density functional theory are $V_{pp\pi} = -2.81~\text{eV}, V_{pp\sigma} = 0.48~\text{eV},
a_\pi = 1.418~\text{\AA},  a_\sigma = 3.349~\text{\AA},
q_\pi = 3.145,  q_\sigma = 7.428, r_c = 6.14~\text{\AA}, l_c = 0.265~\text{\AA}.$ Beyond nearest layer couplings, we include a small second-neighbor interlayer hopping $t_2 \approx -7$ meV, which is essential for reproducing the small intrinsic gap of trilayers \cite{herzog-arbeitman_moire_2024}.

Two additional single particle terms account for experimentally relevant symmetry breaking effects. An inversion symmetric potential (ISP) distinguishes outer and inner layers,
\begin{equation}
[H_{\text{ISP}}]_{ll'} = V_{\text{ISP}} \, \delta_{ll'} \,\Big|\, l - \frac{n+1}{2} \,\Big|,
\end{equation}
with $V_{\text{ISP}}/d_0 \approx 5$ meV/\AA, where $d_0$ is the interlayer spacing of adjacent carbon layers. 
A perpendicular displacement field is modeled as a linear layer dependent potential,
\begin{equation}
[H_D]_{l\alpha,l'\beta} = U \left(l - \frac{n-1}{2}\right) \delta_{ll'}\delta_{\alpha\beta},
\end{equation}
with $U = |e| d_0 D/\varepsilon_r$. Together, these terms capture the tunable interlayer asymmetry and gap opening under external fields. This SK Hamiltonian provides a microscopic foundation for our continuum modeling and phase diagram analysis of RMG.

When the displacement field induced onsite potential satisfies $U>0$, the conduction band electrons become predominantly localized on the $B_N$ orbital, while the valence-band electrons localize mainly on the $A_1$ orbital. Here, \(A_1\) and \(B_N\) label the outermost sublattice orbitals on the top and bottom surfaces of rhombohedral multilayer graphene. Because the low energy states near charge neutrality reside mainly on these two surface orbitals, the problem can be reduced to an effective two band description in the \(A_1\)--\(B_N\) subspace. For $U<0$, the situation is reversed. It is worth noting that, in the bulk limit and in the absence of a displacement field, rhombohedral multilayer graphene is a nodal line semimetal whose geometric configuration closely resembles a three dimensional SSH model \cite{lau_Designing_2021_prx}. The lowest energy surface bands in this limit correspond to the characteristic drumhead states associated with the nodal line topology \cite{chen_surface_2019_prl,lau_Designing_2021_prx}. A representative single particle band structure near the atomic $K$ valley is shown in Fig.~\ref{fig1}(a) for $U = 0.035~\text{eV}$. In this case, the conduction band minimum is remarkably flat, and the associated van Hove singularity lies in the same energy window. Note that these bands are nearly fourfold degenerate due to the spin and valley degrees of freedom. Consequently, one may anticipate a Stoner instability once the criterion $\rho(E) u > 1$ is satisfied, where $\rho(E)$ is the density of states at the Fermi level $E$ and $u$ is the effective Hubbard interaction \cite{zhou_half-_2021}.

\begin{figure}[t]
\includegraphics[width=0.45\textwidth]{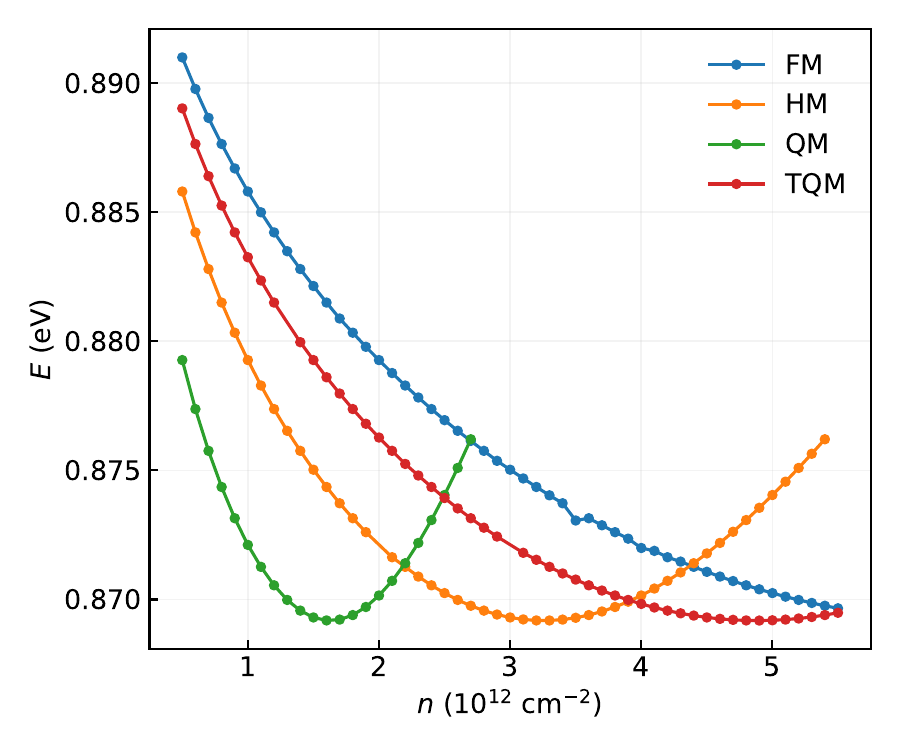}
\caption{\label{fig2}
Cascade of isospin phase transitions in R5G. 
QM denotes the quarter metal phase, HM the half metal phase, TQM the three quarter metal phase, 
and FM the full metal phase. The Hartree Fock results indicate a first order phase transition as the carrier density $n$ increases.
}
\end{figure}

\begin{figure*}[t]
\includegraphics[width=\textwidth]{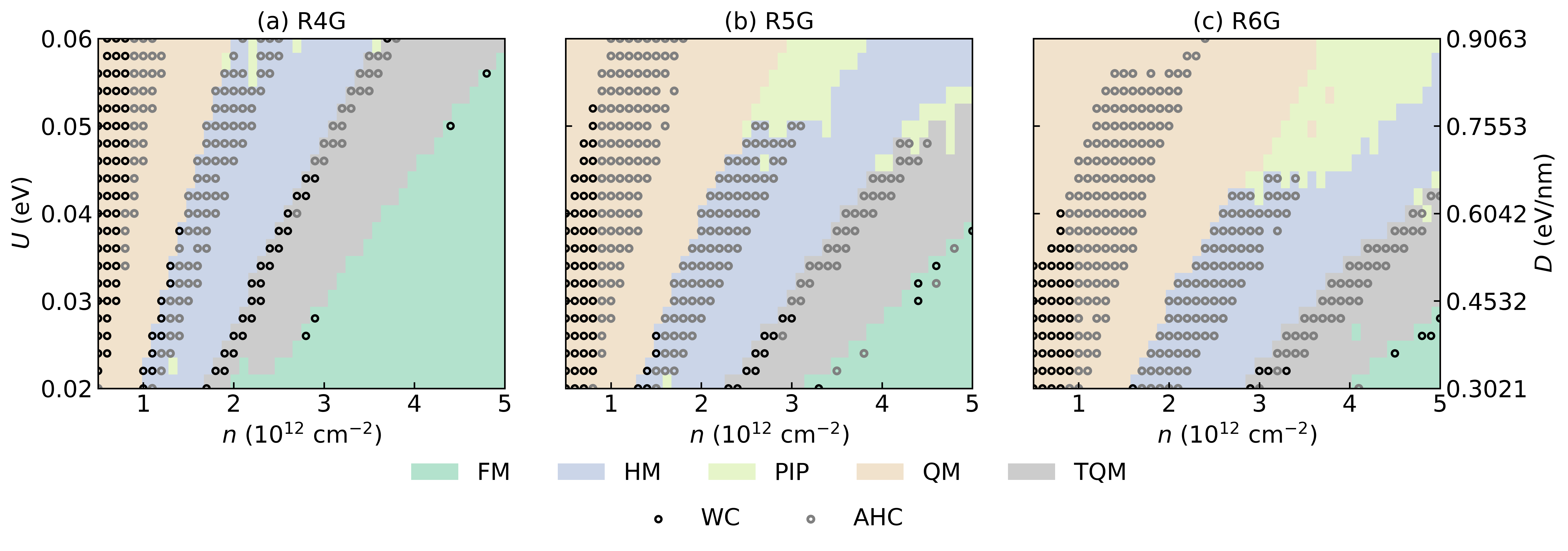}
\caption{\label{fig3} Hartree–Fock mean-field phase diagrams of rhombohedral graphene multilayers as functions of electron doping density 
$n$ and displacement field 
$U$: (a) tetralayer (R4G), (b) pentalayer (R5G), and (c) hexalayer (R6G). QM, HM, TQM, and FM denote the quarter-metal, half-metal, three-quarter-metal, and full-metal phases, respectively. PIP indicates a partially isospin polarized state. Across the isospin driven phase transitions, symmetry-broken electron crystal phases, Wigner crystal (WC) and anomalous Hall crystal (AHC), can also emerge. The annotated regions additionally label the number of isolated flat bands participating at the Fermi level: 1 in QM, 2 in HM, 3 in TQM, and 4 in FM.}
\end{figure*}

\subsection{Mean field treatment}

To capture the Coulomb interaction in the low doping regime, we retain its long range component and neglect the short range part  \cite{koh_correlated_2024,koh_symmetry-broken_2024, guo_fractional_2024, liu_layer-dependent_2025} to perform self consistent Hartree Fock calculations. The interacting part is written as,
\begin{equation}
H_{\mathrm{int}}
=\frac{1}{2}\sum_{\mathbf q} v(\mathbf q)\,:\rho_{\mathbf q}\rho_{-\mathbf q}:\, ,
\label{eq:A1}
\end{equation}
The density operator written in the plane wave basis is
\begin{equation}
\rho_{\mathbf q}
=\sum_{\mathbf k}\sum_{s\mu\alpha}
\,c^\dagger_{\mathbf k+\mathbf q, s\mu\alpha}\,c_{\mathbf k,s\mu\alpha}.
\end{equation}
where $s$ is the spin index, $\mu$ is the valley index and $\alpha$ is the joint index for sublattice ($\sigma$) and layer ($l$). The screened interaction is taken as \cite{kwan_moire_2025},
\begin{equation}
v_{ll'}(\mathbf q)=\frac{e^{2}}{2S\,\epsilon_r \epsilon_0\, q}\times
\begin{cases}
\tanh(qd_s), & l=l',\\
e^{-q|l-l'|d_0}, & l\neq l'.
\end{cases}
\end{equation}
where $d_s = 10$ nm is the gate distance, $d_0$ is the interlayer distance of graphene layers, $S$ is the total area and $\epsilon_r  = 5$ is the dielectric constant.
For a fixed carrier density $n$, the real-space lattice constant (and hence the superlattice length scale) is uniquely determined: $L_s=\sqrt{2/\sqrt{3} n}$ for the hexagonal lattice and $L_s=\sqrt{1/n}$, for the square lattice. We initialize the single particle density matrix by randomly distributing electrons among the spin valley sectors and iteratively solve the Hartree Fock Dyson equation to self-consistency. The zero temperature ground state is then identified as the converged solution with the lowest condensation energy among competing phases. Further computational details are provided in Appendix~\ref{app:hf}, and two recent reviews of Hartree Fock formalism for moir\'e materials are also presented in Ref.~\cite{kwan_mean-field_2025,Lu_numerical_2025}.

\section{electron crystal Phases}

\subsection{Isospin phase transition and full phase diagram}
When the first conduction band of rhombohedral multilayer graphene (RMG) is electron doped toward the van Hove singularity (VHS), a Stoner instability is expected to occur, potentially generating a cascade of isospin phase transitions ~\cite{dong_isospin-_2023,huang_competition_2025,herasymchuk_correlated_2025,Parra-Martinez2025-bj}. Such a cascade has already been explored experimentally in rhombohedral trilayer graphene (R3G)~\cite{zhou_half-_2021}. In our HF calculations, the isospin transitions in RMG appear as first order transitions at the mean field level, as dipicted in Fig.~\ref{fig2}.

A natural question is whether, near these isospin transitions, translational symmetry breaking electron crystal phases can simultaneously emerge and even become energetically favorable compared with the corresponding isospin polarized Fermi liquids. To address this, we map out the full phase diagram of RMG, including the isospin transitions, and highlight the parameter regions where electron crystal phases are the mean field ground states (Fig.~\ref{fig3}). The geometry for these electronic phases is fixed as hexagonal. 

\begin{figure}[t]
\includegraphics[width=0.5\textwidth]{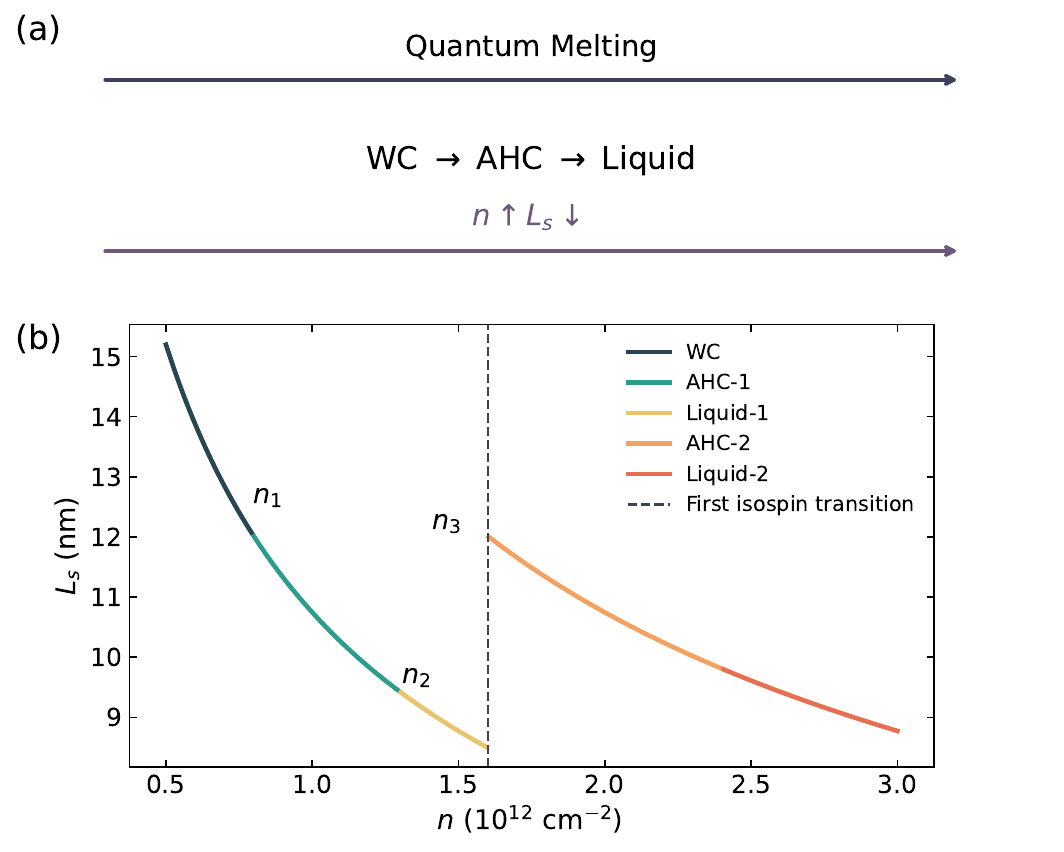}
\caption{\label{melt} 
(a) Schematic illustration of the density driven quantum melting process, in which the anomalous Hall crystal (AHC) serves as an intermediate phase between the Wigner crystal (WC) and the liquid phase. (b) Average electron spacing \(L_s\) within each spin-valley sector as a function of the total doping density \(n\). Here, \(n_1\) marks the WC--AHC transition, \(n_2\) marks the AHC--liquid transition, and \(n_3\) marks the isospin transition. The values of \(n_1\), \(n_2\), and \(n_3\) shown in this figure are extracted from the R5G phase diagram at \(U=0.04~\mathrm{eV}\).
}
\end{figure}

Several general features emerge from the phase diagram. In the low-doping regime, where the system remains fully spin valley polarized, we find a characteristic sequence of phases as the carrier density increases. At the lowest densities, the ground state is a conventional Wigner crystal (WC), reflecting the strong real-space localization favored by Coulomb repulsion. Beyond a critical density, this WC evolves into an anomalous Hall crystal (AHC), and upon further doping the crystalline order eventually melts into a Hall liquid. This density-driven evolution is highlighted in Fig.~\ref{melt} and suggests that the AHC is an intermediate phase in the quantum melting process.

A natural physical picture is as follows. At very low carrier density, electrons are strongly localized and form a nearly classical WC to minimize their electrostatic energy. As the density increases, the charge distribution becomes less rigid, allowing the system to develop a topological crystalline state with reduced localization, namely the AHC, before finally entering an itinerant Hall liquid phase at higher density.

Another notable feature is the ``refreezing'' that occurs at the first isospin phase transition, near density \(n_3\). Across this transition, carriers redistribute among additional flavor sectors. As a result, the carrier density within each individual flavor is abruptly reduced, and the corresponding average intraflavor particle spacing \(L_s\) increases. For the first isospin transition, this redistribution can drive the effective density in a given flavor back below the melting threshold \(n_2\). In our calculations, this typically occurs in the regime \(n_1 \lesssim n_3/2 \lesssim n_2\). Consequently, although the total density increases across the isospin transition, the reduced density per flavor favors the reappearance of crystalline order, leading to a refreezing into the AHC state. These electron crystal phases emerging in the half-metal regime have also been discussed recently~\cite{kudo_quantum_2024_prb, uchida_non-abelian_2025}. In our calculations for the case of two electrons per crystal, when only the long range Coulomb interaction is retained, several competing topological states (spin polarized, valley polarized and spin valley locked) can be nearly degenerate due to the $\mathrm{SU}(2)_{-}\times \mathrm{SU(2)_{+}}$ symmetry \cite{chatterjee_inter-valley_2022}. Additional short range interaction (for example, intervalley Hund's coupling) and intrinsic spin orbital coupling \cite{chatterjee_inter-valley_2022, koh_correlated_2024,koh_symmetry-broken_2024,liu_layer-dependent_2025}) can lift this degeneracy and favor either a fully spin polarized state or a spin valley locked state. Furthermore, this can also explain the absence of triple quarter metal (TQM) phases in experiments \cite{zhou_half-_2021,Auerbach_isospin_2025}. We note that, in the density window \(n \approx(4\!-\!5)\times10^{12}\,\mathrm{cm^{-2}}\), these electron crystal states remain higher in energy than the fully symmetric metallic phase (FM) in our calculations.

We also find a systematic trend with increasing layer number: the electron crystal phases tend to shift toward higher doping densities. At the same time, the density window over which these phases remain stable becomes narrower at large displacement field for thicker rhombohedral multilayers. This behavior can be understood from the evolution of the single particle density of states. In particular, for hexalayer case the van Hove singularity is more strongly suppressed by a large displacement field than the pentalayer, which reduces the tendency toward interaction driven electron crystallization. Details of the corresponding density of states calculations are provided in the Appendix.~\ref{app:dft}.

\subsection{Competing phases and pressure effects in the spin valley polarized case}

For the tetralayer and pentalayer cases, we find that the spin valley polarized anomalous Hall crystal (AHC) phases persist over a relatively wide density range at suitable displacement fields. Recent experiments in R5G/hBN reported an extended quantum anomalous Hall (EQAH) regime: a robust
$R_{xy}\!\approx\! h/e^{2}$ plateau with strongly suppressed $R_{xx}$ persisting over a wide density window,
often accompanied by hysteresis and a pronounced sensitivity to temperature and bias current
\cite{lu_fractional_2024, lu_extended_2025}.
This phenomenology suggests that EQAH is not a single rigid commensurate phase and a deeper understanding for such phenomena is needed.

\begin{figure}[t]
\includegraphics[width=0.48\textwidth]{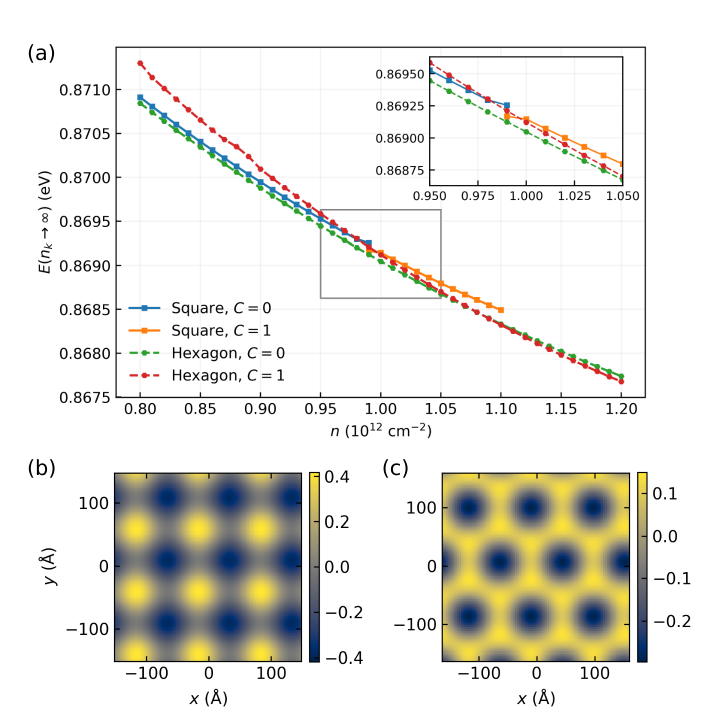}
\caption{\label{fig4}
Competition among different electron crystal phases in R5G as the electron density $n$ varies, with the displacement field fixed at $U = 0.035~\text{eV}$. 
(a) The energy differences between the square lattice states ($C=0/1$) and the hexagonal lattice states ($C=0/1$) are relatively small. 
In certain density ranges, the square lattice phases compete closely with the hexagonal ones. Density profile $(\rho(r)-\left<\rho(r)\right>)\Omega_0$ when $n=1\times 10^{12}$ cm$^{-2}$ for (b) square AHC (c) hexagonal AHC where the changes resemble a honeycomb unit cell. $\Omega_0$ is the superlattice size.}

\end{figure}

Motivated by the fact that the electron crystal states typically develop on a length scale of \(10~\mathrm{nm}\), much larger than the atomic lattice constant, we extend our HF calculations to include square lattice crystal geometries for the spin valley polarized case. At fixed \(U=0.035~\mathrm{eV}\), Fig.~\ref{fig4} shows that the square and hexagonal lattice electron crystal states are nearly degenerate in energy over a wide density window around \(n\sim 10^{12}~\mathrm{cm}^{-2}\). In this situation, an extended \(C=1\) plateau can arise because the system may preserve a quantized Hall response while reorganizing among nearly degenerate AHC realizations as \(n\) is tuned, rather than being locked to a unique commensurate state~\cite{lu_extended_2025}. Moreover, the small energy splittings near the WC--AHC boundary naturally promote metastability and hysteresis.

Upon increasing temperature, the AHC phase may melt into a distinct metallic state at half filling, often discussed in terms of a composite Fermi liquid (CFL) with a larger real space periodicity than the AHC. Although we do not consider such finite temperature phases here, related behavior has been reported experimentally in tetralayer and pentalayer systems~\cite{lu_extended_2025}. Another important distinction between the AHC and the WC is the comparatively reduced stiffness of the AHC \cite{dong_phonons_2025,kwan_moire_2025}, which can be traced to the Wannier obstruction \cite{po_fragile_2018_prl} of the underlying Chern band. This difference is directly reflected in the density profiles of the AHC states shown in Fig.~\ref{fig4}(b) and \ref{fig4}(c). The reduced stiffness can further enhance the susceptibility of the AHC to depinning and depinning \cite{patri_extended_2024,zeng_berry_2025,Shao_electrical_2024_prx} under external fields or in the presence of disorder \cite{joy_disorder-induced_2025}. We also note that a patterned gate, capable of manually modulating the electrostatic potential in the graphene layers, may provide an additional experimental handle for controlling the pinning of these electron crystal states \cite{Ghorashi_topological_2023_prl,Miao_artificial_2025_prb,Shi_fractional_2025}.

\begin{figure}[t]
\includegraphics[width=0.48\textwidth]{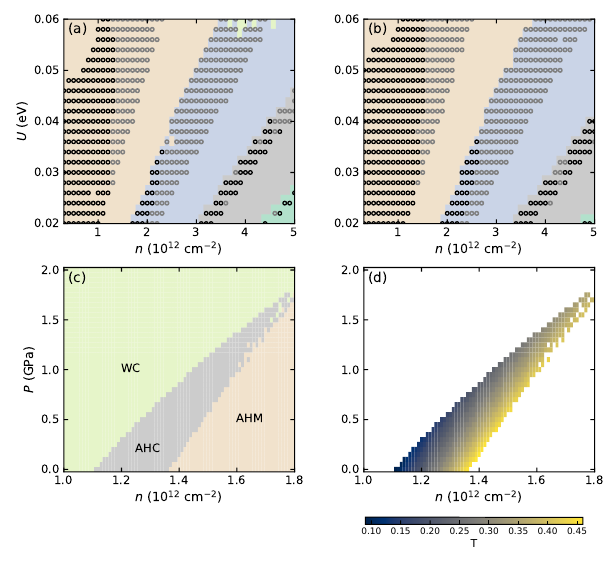}
\caption{\label{fig5}
Pressure induced phase transition in rhombohedral pentalayer graphene (R5G) (a) The same phase diagram as Fig.~\ref{fig3} when $P=0.5$ GPa (b) when $P=1.0$ GPa. The same color legend has been adopted here.
(c) Hartree Fock phase diagram as a function of pressure $P$ and doping density $n$ at fixed displacement field $U=0.035~\mathrm{eV}$. AHM represents the anomalous Hall metal (AHM) region.
(d) Trace condition $T$ corresponding to the anomalous Hall crystal (AHC) phases identified in (c). The dark colored region marks the optimal trace condition, located near the WC--AHC phase transition boundary.}
\end{figure}

To further investigate this near degeneracy, we introduce pressure as an additional tuning knob. In graphene/hBN moir\'e structures, pressure is known to substantially modify the band structure and can enhance the effective moir\'e potential by reducing the interlayer separation \cite{wang_pressure-driven-moire_2025}.
In moir\'e TMDs, pressure has also been proposed as a route to optimize band geometry \cite{morales-duran_pressure-enhanced_2023, wang_pressure-tunable_2025}
to approach an ideal trace condition and thereby influence the competition between charge orders and topological correlated phases.
Here, motivated by these developments and by the EQAH phenomenology \cite{lu_extended_2025}, we examine how pressure reshapes the WC--AHC competition in our model. Importantly, in our calculations pressure is implemented only through the pressure dependent intrinsic band parameters of R5G and we do not include any pressure induced change of the moir\'e potential strength. The interlayer distance and bonding length of carbon atoms is determined through density functional theory, see Appendix \ref{app:dft}. Then the pressure effect on R5G is reflected through the \textit{ab initio} tight binding model described in Sec.~\ref{sec:tb}.
Within this assumption, we first shown the calculated phase diagram for R5G at 0.5 GPa and 1.0 GPa seperately in Fig.~\ref{fig5} (a)-(b). Our results illustrate that the electron crystal states can be further stabilized at higher doping range when the applied pressure increase. Fig.~\ref{fig5}(c) shows that the electron crystal phases extended in the phase diagram while the AHC stable region shrinks
as pressure increases. Nevertheless, Fig.~\ref{fig5}(d) shows near the WC--AHC transition, for moderate pressure $P\lesssim 1.5~\mathrm{GPa}$, the trace condition $T\simeq 0.1$ \cite{dong_anomalous_2024} (dark region in Fig.~\ref{fig5}-(b)). This value is defined as, 
\begin{equation}
    T = \frac{1}{2\pi} \int d \bk (\mathrm{Tr}(g(\bk
    )-|\Omega(\bk)|),
\end{equation}
where $g_{\mu\nu}(\bk)$ is the Fubini-Study metric and $\Omega(\bk)$ is the Berry curvature. This close to zero numeric indicates that the pressure induced suppression of AHC in our calculation is not primarily driven by a degradation of the ideal band geometry condition.

Finally, we emphasize that our analysis is based on static mean field HF energetics and therefore does not include correlation energy beyond HF \cite{guo_correlation_2025,Huo_does_2025}. HF theory is known to overestimate the tendency toward crystallization in the two-dimensional electron gas \cite{guo_correlation_2025}. However, recent variational Monte Carlo studies demonstrate that Berry curvature of the parent band can dramatically enhance crystallization \cite{Valenti_quantum_2025}.
Given the small HF energy differences in Fig.~\ref{fig4}, the correlation effects may be decisive near
the WC--AHC boundary.
Incorporating correlation energy beyond HF, as well as the experimentally relevant pressure-enhanced
moir\'e potential, is therefore an important direction for future work \cite{lu_general_2025,guo_correlation_2025,wang_pressure-driven-moire_2025}.

\section{Discussion}

We discover multiple phase transitions numerically within our mean field treatment of RMG, and we now discuss their thermodynamic signatures in connection with recent experiments \cite{aronson_displacement_2025}. The cascade of isospin phase transitions shown in Fig.~\ref{fig2} is closely related to the stripe-like regions of negative inverse compressibility observed near integer fillings in R5G, where the inverse compressibility is defined as
$K^{-1}(n)\equiv \partial\mu/\partial n$,
with $\mu$ the intrinsic chemical potential of the electronic subsystem.

A useful thermodynamic reference point is a first order transition with macroscopic phase coexistence: the Maxwell construction yields a mixed phase window with constant $\mu$ and hence $K^{-1}=0$ in the thermodynamic limit. In contrast, experiments typically show additional ``shadow'' features in the inverse compressibility, including negative dips, rather than a broad $K^{-1}\!\approx\!0$ plateau. This phenomenology is observed both across the isospin phase transitions and near the metal--insulator boundary. More specifically, Ref.~\cite{aronson_displacement_2025} focuses on the moir\'e proximal region and reports negative inverse compressibility signatures near the quantum freezing region at fractional fillings ($\nu=1/2,\,2/3$ which corresponds to $n\approx (0.4-0.6) \times  10^{12}$ cm$^{-2}$), where these incompressible phases are interpreted as topological charge density waves (TCDWs) with Chern number $C=-1$ at small magnetic field. Notably, this regime overlaps with the electron crystal phases in our calculations.

The sharp negative inverse compressibility \cite{Eisenstein_negative_1992_prl, Eisenstein_compressibility_1994_prb} features cannot be explained by simple macroscopic two-phase coexistence. Instead, long range Coulomb interactions provide a natural route to such behavior: they strongly penalize macroscopic charge separation and can favor intermediate regimes with mesoscopic domain patterns, sometimes discussed as electronic microemulsions \cite{spivak_phases_2004_prb}. Moreover, the resulting phase competition and inhomogeneous coexistence can be further shaped by disorder potentials \cite{joy_disorder-induced_2025, Das_Sarma_thermal_2024_prb}.

In realistic gated devices, further care is required because the experimentally inferred compressibility contains both intrinsic and geometric electrostatic contributions. The total thermodynamic potential includes the classical charging energy associated with transferring charge to and from gate electrodes, which can be recast as a geometric capacitance term \cite{Rai_dynamical_2024_prx}. Consequently, a negative value of \(\partial\mu/\partial n\) for the intrinsic electronic subsystem does not necessarily imply an instability of the full device, since the geometric contribution can restore the stability of the total response. In our Hartree Fock treatment, we impose global neutrality by removing the \(q=0\) Hartree component, equivalently setting \(V(q=0)=0\), so that the remaining interaction captures intrinsic electronic thermodynamics. It is instructive to contrast this with dynamic mean field theory (DMFT) calculations of magic angle twisted bilayer graphene in the heavy fermion language \cite{Rai_dynamical_2024_prx, Datta2023-ur}, where negative inverse compressibility can arise from charge reshuffling between localized and itinerant degrees of freedom upon doping. Once geometric capacitance contributions are included, the negative compressibility regions are often strongly reduced, again emphasizing that negative experimental signals need not correspond to a thermodynamic instability of the full electrostatic environment.

We end this paper by noting that recent reported evidence of chiral superconductivity in R4G, R5G \cite{han_signatures_2025}, and R6G \cite{ron_hierarchy_2025} and bubble-like TCDWs in R6G \cite{ron_hierarchy_2025} at electron dopings of roughly $(0.5\!-\!1.0)\times10^{12}\,\mathrm{cm}^{-2}$ overlap with the topological electron crystal regimes in our mean field phase diagram, suggesting a complex interplay between superconductivity and competing electronic orders. In these experiments the hBN layers are not aligned with the RMGs.

\section{Acknowledgment}

We acknowledge Xi Dai, Binghai Yan and Chunli Huang for stimulating discussion on this topic. W. Miao recognizes support from the Center for Theory of Emergent Quantum Matter (C-TEQ) and Penn State Institute for Computational and Data Sciences (RRID:SCR025154) for providing access to computational research infrastructure within the Roar Core Facility (RRID: SCR026424).

\section{Code \& Data availability}

The Hartree Fock code used in this work is available on GitHub \texttt{https:/github.com/zybbigpy/R5G-AHC} and the data generated using this code is available through request to the authors.

\newpage

\bibliography{reference}

\appendix

\begin{figure*}[h]
\includegraphics[width=0.9\textwidth]{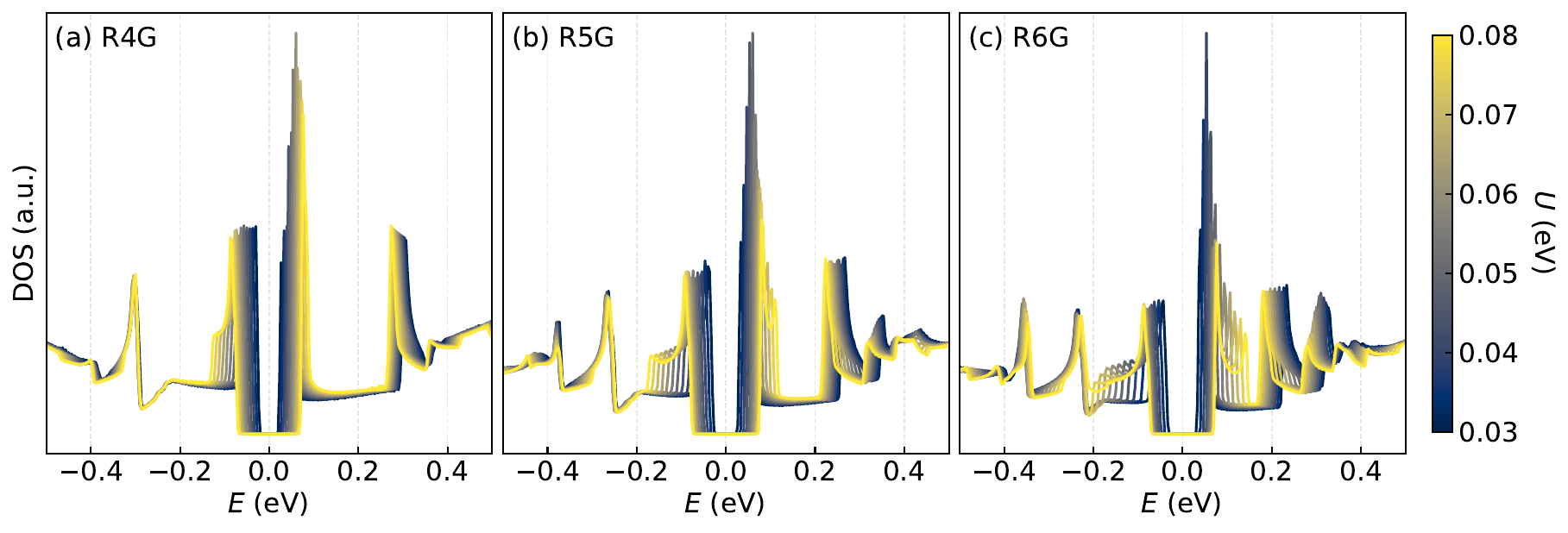}
\caption{\label{fig-dos}
Density of states for pristine (a) R4G (b) R5G (c) R6G at different onsite energy. The stronger displacement field will suppress the van-Hove singularity for R6G compared with R4G and R5G.}
\end{figure*}

\begin{widetext}\section{DFT under pressure and density of states \label{app:dft}}

The density of states for R4G, R5G and R6G with different onsite energy (under different displacement field) is shown in Fig.~\ref{fig-dos}. The von-Hove singularity (VHS) corresponds to the flat band bottom emerged in RMG, while a strong displacement field can suppress VHS especially in the case of R6G.

To determine the interlayer spacing and lattice constant of graphene under pressure, we performed density functional theory (DFT) calculations using the Vienna Ab initio Simulation Package (VASP) \cite{Kresse_efficient_1996_prb}. The exchange–correlation effects were treated within the generalized gradient approximation using the revised Perdew–Burke–Ernzerhof (RPBE) functional \cite{perdew_generalized_1996_prl}, together with projector augmented-wave (PAW) pseudopotentials \cite{Blochl1994-pa}. Van der Waals interactions were included via the DFT-D3 correction scheme \cite{Grimme2010-ee}. Pentalayer graphene was modeled using a five-layer $3\times3$ supercell with ABC stacking. Prior to extracting the interlayer distances, all carbon atomic positions were fully relaxed under the applied external pressure. A plane-wave energy cutoff of 520~eV was used for all pressure-dependent calculations. The total energy and force convergence thresholds were set to $10^{-8}$~eV and $10^{-3}$~eV~\AA$^{-1}$, respectively. The relaxed lattice constant $a$ and interlayer distance $d$ is shown in Fig.~\ref{fig_app0}.

\begin{figure}[t]
\includegraphics[width=0.9\textwidth]{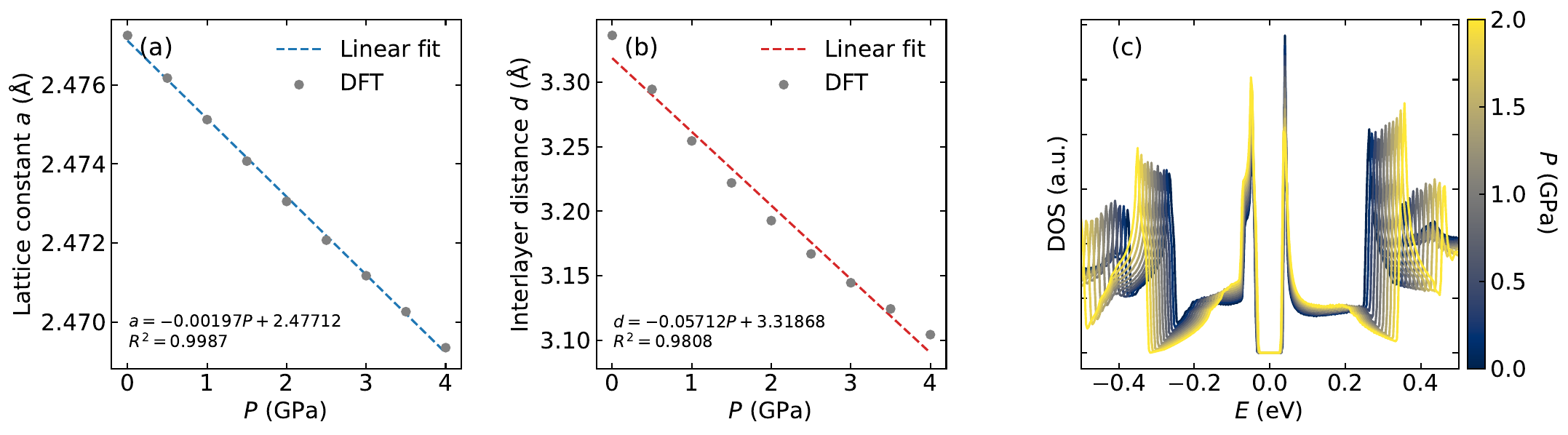}
\caption{\label{fig_app0}
(a) (b) The relaxed lattice constant and interlayer distance of R5G under moderate pressure. The grey dots are extracted from DFT calculations and the linear fit is in use for calculate the phase diagram in Fig.~\ref{fig5}. (c) Calculated density of state for R5G under pressure when displacement field is constrained.}
\end{figure}

\section{Hartree Fock \label{app:hf}}

To perform the Hartree Fock calculations, we first project the interacting Hamiltonian Eq.~\ref{eq:A1} onto active subspace,
\begin{equation}
    H_{\mathrm{int}} =
    \frac{1}{2N_{\bkbar}}
    \sum_{\bar{\mathbf{k}},\bar{\mathbf{k}}',\bar{\mathbf{q}}}
    \sum_{\mu\mu'}\sum_{ss'}\sum_{nm n'm'}
    \left( \sum_{\bQ} v(\bQ+\bar{\mathbf{q}})\,
    \lambda^{\mu\mu'}_{nm,n'm'}(\bar{\mathbf{k}},\bar{\mathbf{k}}',\bar{\mathbf{q}},\bQ) \right)
    d^\dagger_{s\mu n}(\bar{\mathbf{k}}+\bar{\mathbf{q}})
    d^\dagger_{s'\mu'n'}(\bar{\mathbf{k}}'-\bar{\mathbf{q}})
    d_{s'\mu'm'}(\bar{\mathbf{k}}')\,
    d_{s\mu m}(\bar{\mathbf{k}}),
\end{equation}
where $\bkbar, \bkbar', \bar{\bq}$ are momentums defined in the mini Brillouin zone (mBZ) due to band folding,  \(N_{\bkbar}\) is the number of momentum points in the mBZ grid.  $\bG, \bQ$ are reciprocal lattice vectors. The form factor is defined as
\begin{equation}
    \lambda^{\mu \mu' s s'}_{nm,n'm'}(\bar{\mathbf{k}},\bar{\mathbf{k}}',\bar{\mathbf{q}},\bQ)
    = \sum_{\alpha\alpha'\bG \bG'}
    C^{*}_{\mu \alpha s \bG+\bQ,n}(\bar{\mathbf{k}}+\bar{\mathbf{q}})\,
    C^{*}_{\mu' \alpha's' \bG'-\bQ,n'}(\bar{\mathbf{k}}'-\bar{\mathbf{q}})\,
    C_{\mu' \alpha' s'\bG',m'}(\bar{\mathbf{k}}')\,
    C_{\mu \alpha s\bG,m}(\bar{\mathbf{k}}).
\end{equation}
where we take advantage of 
\begin{equation}
    d_{s\mu n}(\bkbar) = \sum_{\alpha \bG} C_{\mu \alpha s \bG, n}(\bkbar) c_{\mu \alpha s}(\bkbar+\bG).
\end{equation}
We now formulate the band-projected Hartree Fock theory in the Green's-function framework, where the mean field corrections naturally appear as static self energy correction in the Dyson equation. In the mBZ, we use composite band indices \(\eta=(s,\mu,n)\), where \(s\) denotes spin, \(\mu\) the valley, and \(n\) the band index. The single particle Green's function is defined as \([G(\bar{\mathbf{k}},\tau)]_{\eta\eta'}=-\langle T_{\tau} c_{\bar{\mathbf{k}},\eta}(\tau)c^{\dagger}_{\bar{\mathbf{k}},\eta'}(0)\rangle\), satisfying the Dyson equation \(G^{-1}(\bar{\mathbf{k}},i\omega_n)=i\omega_n+\mu-H_0(\bar{\mathbf{k}})-\Sigma(\bar{\mathbf{k}},i\omega_n)\), where \(H_0(\bar{\mathbf{k}})\) is the single particle Hamiltonian projected to the chosen band subspace. In the HF approximation, the self-energy is static, \(\Sigma(\bar{\mathbf{k}},i\omega_n)\equiv\Sigma(\bar{\mathbf{k}})\). The single particle density matrix is defined as \(\rho_{\eta\eta'}(\bar{\mathbf{k}})=\langle c^{\dagger}_{\bar{\mathbf{k}},\eta'}c_{\bar{\mathbf{k}},\eta}\rangle\), and the total self energy consists of Hartree and Fock components, \(\Sigma=\Sigma^{H}+\Sigma^{F}\). These are expressed through four-index form factors \(\Lambda\) constructed from the Bloch coefficients of the projected bands. The Hartree and Fock contributions are given by 
\begin{equation}
    [\Sigma^{H}(\bar{\mathbf{k}})]_{\eta\eta'} = \frac{1}{N_{\bkbar}}\sum_{\bar{\mathbf{k}}',\zeta,\zeta',\mathbf{Q}} v(\mathbf{Q})\,\Lambda^{H}_{\eta\eta';\,\zeta\zeta'}(\bar{\mathbf{k}},\bar{\mathbf{k}}';\mathbf{Q})\,\rho_{\zeta'\zeta}(\bar{\mathbf{k}}'), 
\end{equation}

\begin{equation}
    [\Sigma^{F}(\bar{\mathbf{k}})]_{\eta\eta'} = -\frac{1}{N_{\bkbar}}\sum_{\bar{\mathbf{k}}',\zeta,\zeta',\mathbf{Q}} v(\bar{\mathbf{k}}'-\bar{\mathbf{k}}+\mathbf{Q})\,\Lambda^{F}_{\eta\zeta;\,\zeta'\eta'}(\bar{\mathbf{k}},\bar{\mathbf{k}}';\mathbf{Q})\,\rho_{\zeta'\zeta}(\bar{\mathbf{k}}'),
\end{equation}
where \(N_{\bar{k}}\) is the number of momentum points in the mBZ grid. The tensors \(\Lambda^{H/F}\) entering the self-energies are directly related to the four-index projected form factors \(\lambda\): \(\Lambda^{H}_{\eta\eta';\,\zeta\zeta'}(\bar{\mathbf{k}},\bar{\mathbf{k}}';\mathbf{Q})=\lambda^{\mu\mu'}_{n m,\, n' m'}(\bar{\mathbf{k}},\bar{\mathbf{k}}',\bar{\mathbf{q}}=0,\mathbf{Q})\) for the Hartree channel, and \(\Lambda^{F}_{\eta\zeta;\,\zeta'\eta'}(\bar{\mathbf{k}},\bar{\mathbf{k}}';\mathbf{Q})=\lambda^{\mu\mu'}_{n m,\, n' m'}(\bar{\mathbf{k}},\bar{\mathbf{k}}',\bar{\mathbf{q}}=\bar{\mathbf{k}}'-\bar{\mathbf{k}},\mathbf{Q})\) for the Fock channel, with composite indices \(\eta=(s,\mu,n)\), \(\eta'=(s,\mu,m)\), \(\zeta=(s',\mu',m')\), and \(\zeta'=(s',\mu',n')\). The self-consistent HF equations are then solved in the Dyson framework: starting from an initial random single particle density matrix across different spin-valley flavors. The Hartree and Fock self-energies are evaluated via the above four-index contractions, forming the static Dyson Hamiltonian \(H_{\mathrm{HF}}=H_0+\Sigma^{H}+\Sigma^{F}\), the Dyson equation is solved to obtain an updated Green's function and density matrix, and the process is iterated with mixing until both the density matrix and self-energy converge. At convergence, \(\Sigma(\bar{\mathbf{k}})\) defines the quasiparticle spectrum of RMG electrons. The total energy  is calculated through,

\begin{equation}
    E= \sum_{\bkbar} \mathrm{Tr}\left((H(\bkbar)+\frac{1}{2}\Sigma(\bkbar
    ))\rho(\bkbar
    )\right).
\end{equation}
The real space density $\rho(\mathbf{r})$ can also be obtained,
\begin{equation}
    \rho(\mathbf{r}) = \sum_\bG \rho(\bG) e^{i \bG \cdot\mathbf{r}}, \quad
    \rho(\bG) = \frac{1}{N_{\bkbar}^2} \sum_{\bar{\bk} mn} \braket{u_m^{\mathrm{HF}}(\bar{\bk})|u_n^{\mathrm{HF}}({\bar{\bk}+\bG})}
\end{equation}
where $\ket{u_n^{\mathrm{HF}}}$ is the wavefunction of the $n$-th converged occupied Hartree-Fock band. In all calculations throughout this paper, we only preserve 3 conduction bands, 4 shells of plane waves and freeze all valence bands of RMG due to the relatively large displacement field we concern. Typical Hartree Fock quasi-particle band structures for these electronic phases are shown in Fig.~\ref{fig_app1} with different doping densities and real space geometry. The representative Wilson loop and density profile for the AHC at $n=2\times10^{12}\,\mathrm{cm}^{-2}$ is shown in Fig.~\ref{fig_app3}.

\begin{figure}[t]
\includegraphics[width=0.95\textwidth]{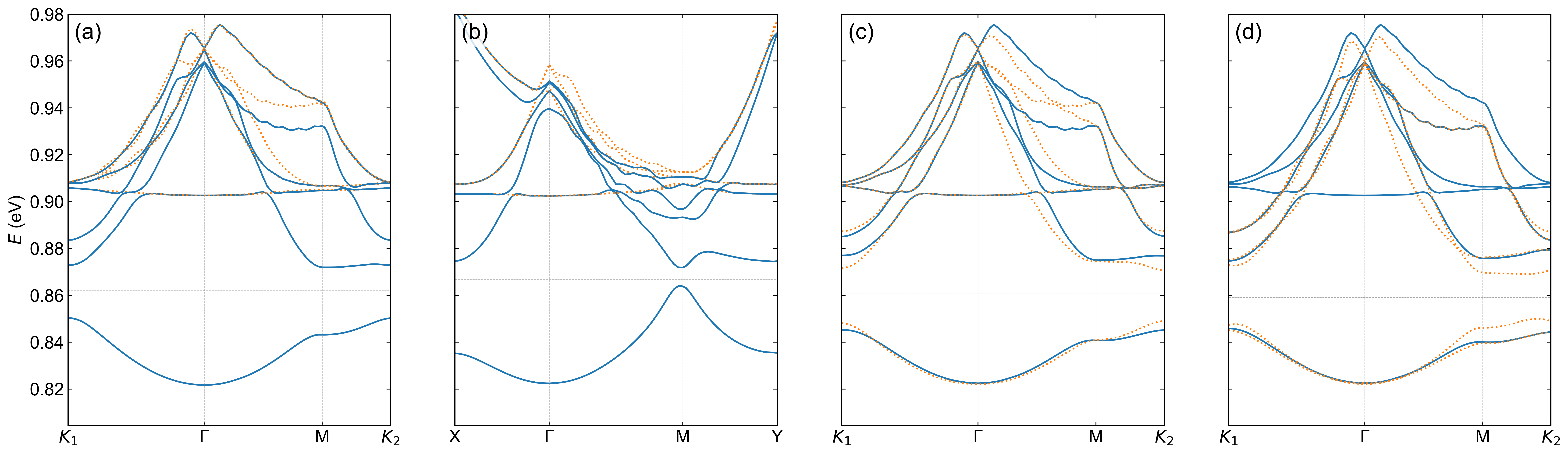}
\caption{\label{fig_app1}
Hartree Fock band structures for representative anomalous Hall crystal (AHC) states. Blue solid lines denote spin–down bands and orange dashed lines denote spin–up bands. 
(a) AHC ($C=1$) on a hexagonal lattice at $n=1\times10^{12}\,\mathrm{cm}^{-2}$; one occupied (or low-energy) band is topological while the other is topologically trivial. 
(b) AHC ($C=1$) on a square lattice at $n=1\times10^{12}\,\mathrm{cm}^{-2}$; one band is topological and the other is trivial. 
(c) AHC ($C=1$) on a hexagonal lattice at $n=2\times10^{12}\,\mathrm{cm}^{-2}$; one band is topological and the other is trivial. 
(d) AHC ($C=1$) on a hexagonal lattice at $n=3\times10^{12}\,\mathrm{cm}^{-2}$; one band is topological and two bands are trivial.
}
\end{figure}

\begin{figure}[t]
\includegraphics[width=0.95\textwidth]{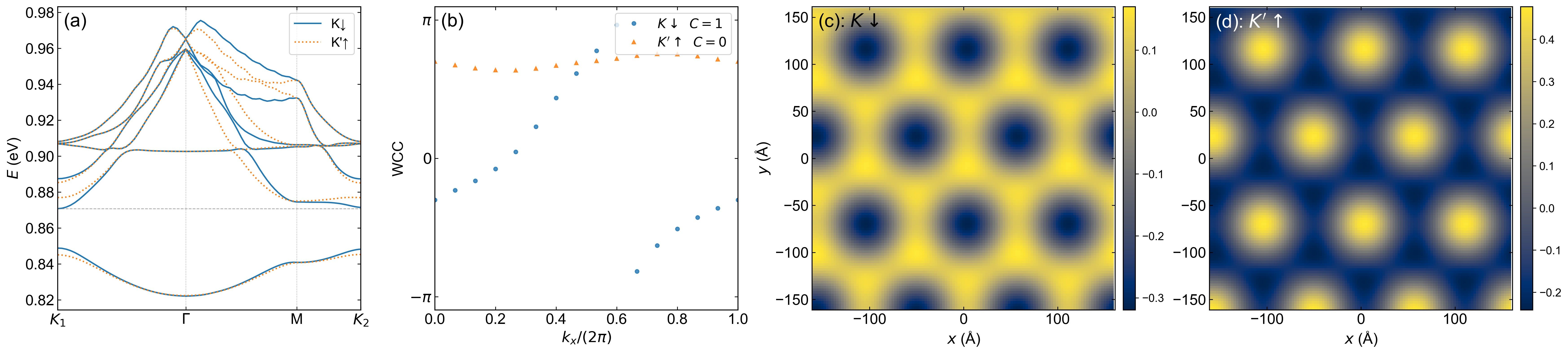}
\caption{\label{fig_app3}
Hartree Fock results for the $n=2\times10^{12}\,\mathrm{cm}^{-2}$, with active flavors $K\downarrow$ and $K'\uparrow$. (a) Band structure. Blue solid (orange dotted) lines denote the $K\downarrow$ ($K'\uparrow$) sector; the grey dashed line marks the chemical potential $\mu$. (b) Wilson loop (Wannier Charge Center) for the lowest occupied band in each sector as a function of $k_x$. The winding numbers give the Chern numbers $C(K\downarrow)$ and $C(K'\uparrow)$ indicated in the legend. (c,d) Real space charge density $\rho(\mathbf{r})\Omega_0 - \langle\rho\rangle\Omega_0$ for the $K\downarrow$ and $K'\uparrow$ sectors, respectively.
}
\end{figure}

\section{Finite size scaling}

To minimize finite size effects and enable a controlled comparison of mean-field condensation energies
among electron crystal states with different lattice geometries, we perform self-consistent
HF calculations on a sequence of increasingly dense $(n_k,n_k)$ momentum meshes.
For each mesh we evaluate the condensation energy per particle, $E(n_k)$.
Since the dominant discretization error is expected to scale with the inverse number of sampled
$k$ points, $N_k=n_k^2$, we extrapolate to the thermodynamic  limit by fitting
$E(n_k)$ linearly in $1/n_k^2$, $E(n_k)=E(n_k \rightarrow \infty)+\frac{a}{n_k^2}$.
The intercept $E(n_k \rightarrow\infty)$ is taken as the infinite $k$-mesh estimate, and the same
extrapolation protocol is applied to all candidate geometries to ensure an unbiased energy comparison. The results is shown in Fig.~\ref{fig_app2}.

\begin{figure}[t]
\includegraphics[width=0.95\textwidth]{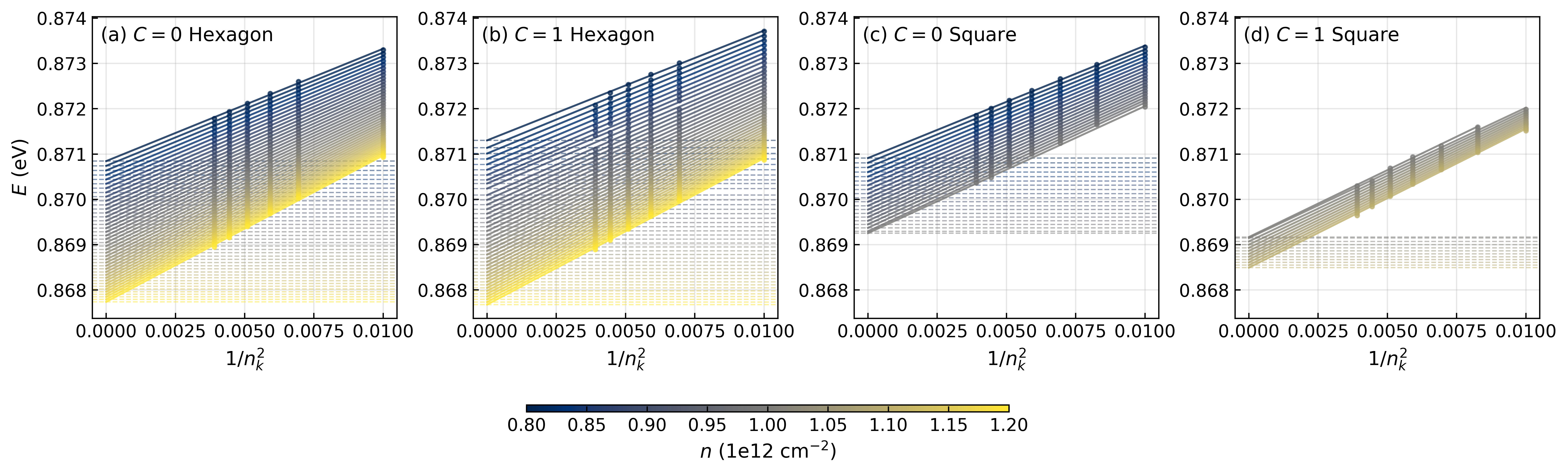}
\caption{\label{fig_app2}
Finite size scaling of the HF condensation energy for electron crystal states with different lattice
geometries. The condensation energy per particle, $E(n_k)$, is computed on $(n_k,n_k)$
$k$-meshes and linearly extrapolated in $1/n_k^2$ to the $n_k\!\to\!\infty$ limit. The intercept
$E(n_k \rightarrow \infty)$ is used to produce Fig.~\ref{fig4} in the main text.
}
\end{figure}

\end{widetext}

\end{document}